\documentclass[12pt]{article}
\usepackage[dvips]{graphicx}
\begin{document}

\date{\today}
\title{Multiple Scales in the Fine Structure of the Isoscalar Giant 
Quadrupole Resonance in $^{208}$Pb}

\author{
D. Lacroix,$^{1}$ 
A. Mai,$^{2}$
P. von Neumann-Cosel,$^{2}$
A. Richter,$^{2}$
and 
J. Wambach$^{2}$ \\
%%\address{
$^{1}$ LPC/ISMRA, Blvd du Mar\'{e}chal Juin, 14050\ Caen, France,\\
$^{2}$ Institut f\"ur Kernphysik, Technische Universit\"at 
Darmstadt, D-64289 Darmstadt, Germany }
\maketitle

\begin{abstract}
The fine structure of the isoscalar giant quadrupole resonance 
in $^{208}$Pb, observed in high-resolution (p,p$'$) and 
(e,e$'$) experiments,
is studied using the entropy index method. In a novel way, it
enables to determine the number of scales present in the spectra 
and their magnitude. 
We find intermediate scales of fluctuations around 1.1 MeV, 460 keV 
and 125 keV for an excitation energy region $0 - 12$ MeV. 
A comparison with scales extracted from second RPA calculations, 
which are in good agreement with experiment, 
shows that they arise from the 
internal mixing of collective motion with 
two particle-two hole components of the nuclear wavefunction.
\end{abstract}



%%\pacs{24.30.Cz, 21.10.Re, 21.60.Jz, 27.80.+w} 

%\vskip2pc
%]

\section{Introduction}

The decay of giant resonances in nuclei provides important
information on how a well-ordered collective excitation 
dissolves into a disordered motion of internal degrees 
of freedom in fermionic quantum many-body systems 
(see e.g.~\cite{BBB98}).
This can be understood to result from the decay of the
collective modes towards compound nuclear states leading to 
internal mixing. 
Besides this internal mixing, the nuclear states
may also decay into a continuum of escaping states 
giving rise to the deexcitation of the system 
through particle emission. 

It is generally agreed that internal mixing
occurs through a hierarchy of couplings towards more and more 
complex degrees of freedom in the nucleus. 
Collective states are constructed in mean-field theory
as a coherent superposition
of one particle-one hole (1p-1h) excitations and are generally treated 
in the random-phase approximation (RPA) \cite{Row70,Rin81}. 
In order to understand their decay towards the compound nucleus, 
one has to go beyond the mean field and introduce two-body
effects as embodied in the second RPA (SRPA) \cite{Dro90}. 
Such approaches are valid under the assumption that collective 
motion is preferentially damped by 2p-2h components of the many-body
wavefunction, reflecting the two-body nature of the nuclear interaction.
The description can be extended by introducing more and more 
complex components such as 3p-3h...np-nh. 
Indeed, all transport theories assume a classification in increasing
degrees of complexity \cite{Cas90,Rei94,Abe96}.

Since this picture is based on a particular hierarchy, 
one should be able to extract experimental information on it by studying 
scales present in the decay of collective motion. 
One expects a hierarchy of lifetimes linked to
a hierarchy of energy scales starting from
the typical scale associated with collective states, the full width at 
half maximum (FWHM) which is of the order of a few MeV, 
going down to scales characterized by the width 
of long-lived compound nuclear states which is of the order 
of a few eV.
%However, the general picture of nuclear transport models
%might well break down at a given level and one 
%might be forced to consider the full complexity of nuclei\cite{Zel96}.  
In order to test this framework, an experimental identification 
of scales involved in the decay of giant resonances appears as 
an important issue.

Experimental evidence for scales associated with the coupling between
collective states and internal and external degrees of freedom is a 
long-standing problem. On the one hand, the spectral analysis 
requires high-resolution experiments. 
Proton and electron scattering experiments, which may reach
resolutions better than 50 keV, present promising candidates 
for this type of analysis. 
The appearance of fine structure in the isoscalar giant
quadrupole resonance (ISGQR) of $^{208}$Pb has been 
reported already a long time ago in high-resolution electron
scattering \cite{Sch75,Kuh81}.
This finding, which has led to considerable debate, was
finally confirmed \cite{Kam97} when proton scattering 
data of comparable resolution became available \cite{Lis91}.
On the other hand, one has to develop tools to extract the
information from a complex signal where several scales of 
different nature are mixed. 
A variety of methods has been proposed to study 
fluctuation properties of the experimental spectra
either using a doorway model and microscopic 
calculations \cite{Win83} or 
taking advantage of autocorrelation 
techniques \cite{Eri63} assuming a statistical 
distribution of the decay channels \cite{Kil87}.
However, such analyses remain dependent on 
underlying model hypotheses and become difficult to 
handle when more than one scale of the fine structure 
exists.

In the present paper, we reanalyze the $^{208}$Pb(p,p$'$) and 
$^{208}$Pb(e,e$'$) experiments using a model-independent method, 
the entropy index, which is especially suited for the 
study of multiscale fluctuations \cite{Lac99}. 
In particular, this method does not make any a priori 
assumptions on the decay mechanism. 
In the next section the entropy index method is briefly summarized. 
Its application is then illustrated for the $^{208}$Pb(p,p$'$)
experiment where fine structure at different scales is found. 
Results of electron and proton scattering are then 
compared showing a perfect agreement between these two independent
measurements.
We finally analyze the results of SRPA calculations of the 
ISGQR in $^{208}$Pb in order to understand the origin of the 
fluctuations. 
A good agreement of scales extracted from the 
calculated and experimental results is found.
This supports the interpretation that the two-body nature of
the nuclear interaction governs the dominant decay channels.

\section{The entropy index method}

The principle of the entropy index method is to have a measure of the 
fluctuations at a given resolution in energy $\delta E$. 
Suppose that we have an experimental spectrum in 
the excitation energy region 
$\Delta E = \left[E_{min},E_{max}\right]$. 
In order to study fluctuations, we can divide this interval 
into $n$ bins with $n=\Delta E/\delta E$. 
If we call $\sigma \left( E\right)$ the
fluctuating function, i.e., the cross section or the strength, 
we define a coefficient $D_j$ in each bin $j$ as            
\begin{equation}
D_j\left( \delta E\right) =\int_{E_{j-1}}^{E_j}dE~\sigma \left( E\right)
\Omega_j(E) 
\label{dj}
\end{equation}
where $E_{j}= E_{min}+j \delta E$. In this expression, $\Omega_j(E)$ 
denotes a function which takes a non-zero value in the interval 
$\left[E_{j-1},E_j \right]$.
In the following, we 
suppose\footnote{It should however be noted that the choice
of $\Omega_j(E)$ is not unique and other odd functions
with respect to the center of the interval could be used.}
that $\Omega_j(E)= {\rm sign}\left( E-\left(j-1/2\right) \delta E\right)$.
With this definition, $D_j$ represents a coarse-grained derivative 
of the function $\sigma$, and the fluctuations of these coefficients
are directly related to the fluctuations of $\sigma$ at the
considered scale. In order to infer global properties of these 
fluctuations, we can define an entropy $K$
\begin{equation}
K\left( \delta E\right) = - \frac {1}{n} \sum_{j=1,n}W_j\left( \delta E\right) 
\log
W_j\left( \delta E\right)
\end{equation}
where $W_j\left( \delta E\right) =$ $\left|D_j \right| /\left\langle
\left|D_j \right| \right\rangle $ stands for the absolute 
value of the coefficients $D_j$ normalized to
their averaged value ($\left\langle \left| D_j \right|
\right\rangle =1/n\sum_{j=1,n}\left| D_j \right|$).
      
This technique has been initially proposed in order to study 
self-similar fluctuations in heartbeats which could be 
identified through a linear dependence
of $K\left( \delta E\right)$ on the logarithm of the 
resolution $\delta E$ \cite{Hwa98}. 
It has recently been shown that the entropy index method is also 
suitable in situations where well-separated scales of 
fluctuations exist \cite{Lac99}. Then, the linear increase is replaced 
by a change in curvature of the entropy which corresponds to transitions
of $\delta E$ from one scale to another.
Note finally that, in order to avoid problems due to the limited number 
of bins for large $\delta E$, in the results presented below 
we have considered a function defined as 31 repetitions 
of the analyzed spectra as described in Ref.~\cite{Lac99}.

\section{Results}

\begin{figure} [h]
\begin{center}
%\epsfig{figure=eim-fig1.ps,
%height=6.0cm,angle=0}%,bbllx=1cm,bblly=3.5cm,bburx=19.5cm,bbury=26cm,clip=}
%\vspace{0.2cm}
\includegraphics*[height=6cm]{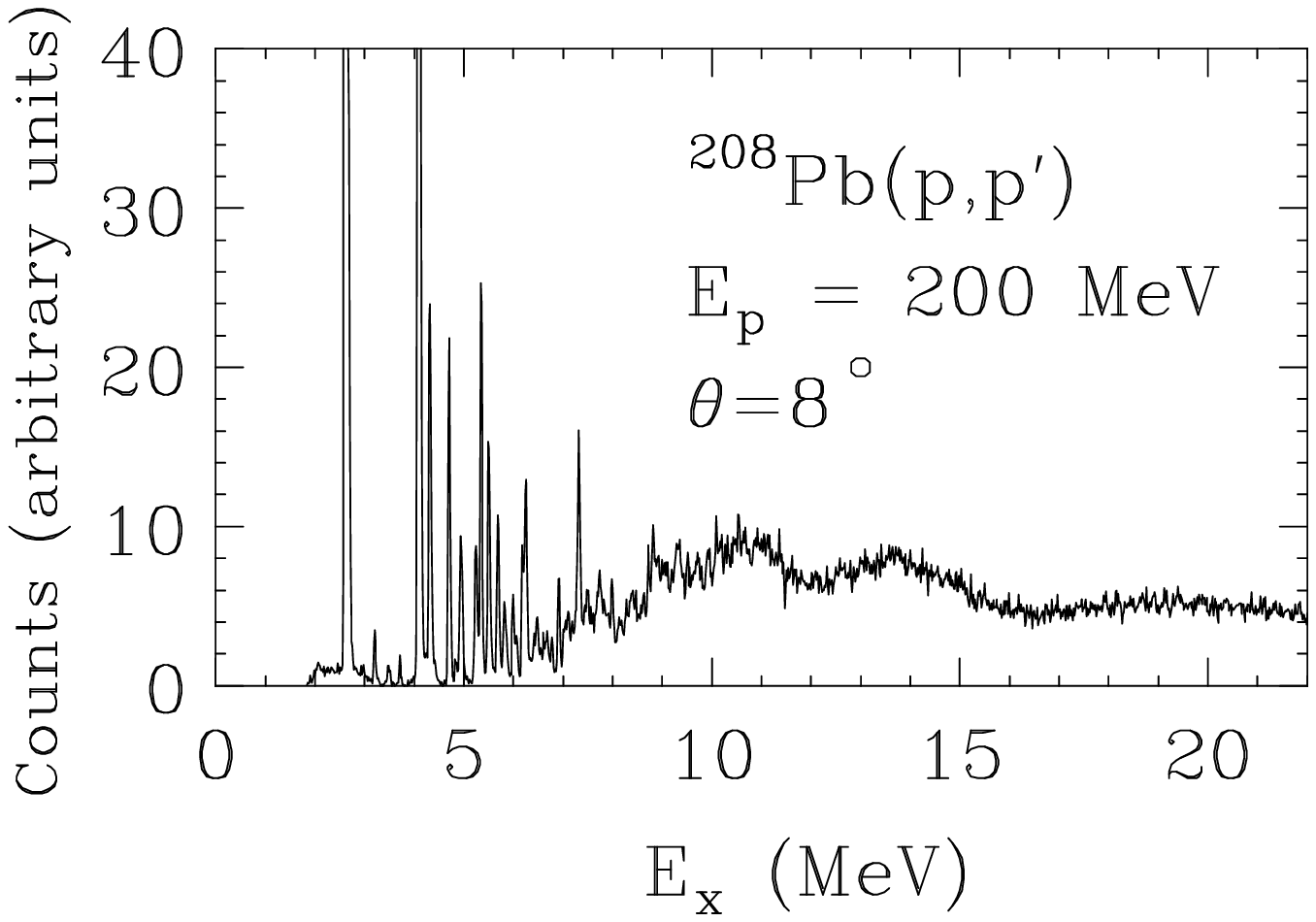}
\end{center}                                     
\caption{The $^{208}$Pb(p,p$'$) experimental spectrum at 
a beam energy $E_p = 200$ MeV and $\theta=8^\circ$.}
\label{fig:1}
\end{figure}

In the following, we apply the presented method to the 
experimental data. 
A sample spectrum of the $^{208}$Pb(p,p$'$) reaction \cite{Lis91}
is displayed
in Fig.~\ref{fig:1} at $\theta = 8^\circ$, where $\Delta L = 2$
transitions - and thus the population of the ISGQR - are enhanced.
With an experimental resolution of about 50 keV fine structure
has been observed at excitation energies below 12 MeV.
Indeed, in a doorway picture
one expects a large increase of the density of 
states to which collective modes are coupled with increasing 
excitation energy, $E_x$. 
As a result, scales present at lower excitation 
energies might differ from those at higher energies. 

\begin{figure}[h]
\begin{center}
\includegraphics*[height=5cm]{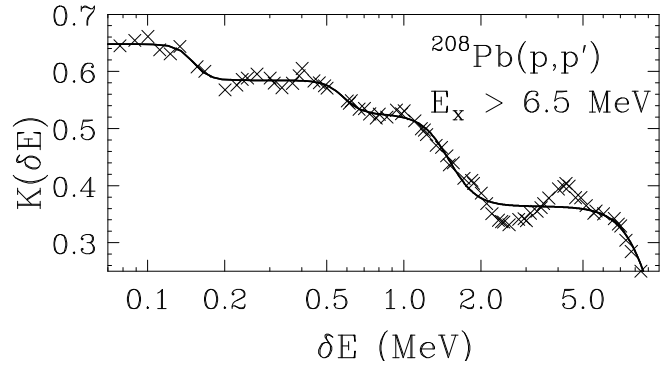}
%\vspace{0.2cm}
\end{center}                                     
\caption{The variation of the entropy index as a function 
of the resolution 
$\delta E$ obtained with the $^{208}$Pb(p,p$'$) data 
when excitation energies $E_x >6.5$ MeV are selected 
(crosses).
The solid line corresponds to a fit of 
Eq.~(\protect\ref{eq:fit})
using a sum of four Fermi-Dirac functions.}
\label{fig:4}
\end{figure}

In Fig. \ref{fig:4}, we show the entropy variation obtained by 
selecting the excitation energy interval $E_x > 6.5$ MeV of the (p,p$'$) 
spectrum displayed in Fig.~\ref{fig:1}. 
We select this interval in order to exclude strongly excited
low-lying levels which are outside the scope of this method. 
Figure \ref{fig:4} exhibits sudden
variations in the evolution of the entropy as a function of $\delta E$
which can be associated with the appearance of different dominant scales 
in the fine structure.  
Although the number of dominant scales appears 
directly from the analysis, their precise determination 
is not straightforward from the localization of the curvature changes.

In order to gain a deeper insight, we have extended the work performed 
in \cite{Lac99}. We have shown that for the models considered in \cite{Lac99} 
the entropy index can be properly fitted by a 
function $F(\delta E)$ defined as $F(\delta E) = \sum_n K_n(\delta E)$ 
where $n$ is an index running on the different scales and where
$K_n(\delta E)$ is defined as a Fermi-Dirac like distribution function   
\begin{equation}
K_n(\delta E)  = \frac{k_n} 
{\displaystyle 1+ 
\exp {\displaystyle \left( \frac{\displaystyle 
\ln(\delta E)- d_n}{\Delta_n} \right)}}
\label{eq:fit}
\end{equation}
with parameters $k_n$, $d_n$ and $\Delta_n$.
An example of the fit obtained with $F$ is shown 
as solid line in Fig. \ref{fig:4}.
In the model investigations \cite{Lac99}, 
where the scales $\Gamma_n$ are known, 
an empirical relation could be established between the 
fitting functions and the scales, viz.\
$K_n(\Gamma_n)/k_n = 0.92 \pm 0.01$. 
Application of the same relation to the experimental data 
allows the identification of fine-structure scales at 1.1 MeV, 
460 keV and 125 keV. 
It should be noted that the value of the smallest scale 
might already be affected by the experimental resolution.           
Furthermore, the procedure described above can only serve 
as an indication of the range of fluctuations since it 
is based on an empirical observation in model cases only.

Considering the larger scales, it is interesting to note that 
assuming a two-doorway picture for the ISGQR in $^{208}$Pb, 
the authors of Ref.~\cite{Win83} 
predicted spreading widths $\Gamma^\downarrow$ of 
490 keV and 740 keV, respectively. 
The first scale in particular 
seems to corroborate our result while the second one is 
somewhat smaller than what is deduced here. 
We would like to emphasize, however, 
that our present method does not suppose any number of doorways.

\begin{figure}[h]
\begin{center}
%\epsfig{figure=eim-fig3.ps,
%height=8.0cm,angle=0}%,bbllx=1cm,bblly=3.5cm,bburx=19.5cm,bbury=26cm,clip=}
%\vspace{0.2cm}
\includegraphics*[height=8cm]{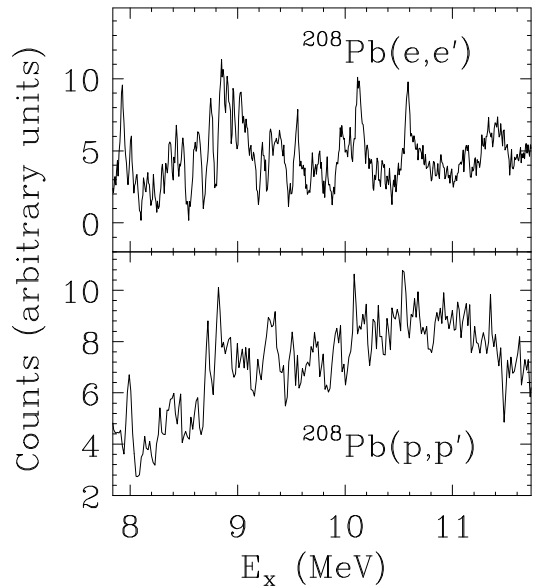}
\end{center}                                     
\caption{Top: the $^{208}$Pb(e,e$'$)
spectrum (where the background is removed) at $E_e = 50$ 
MeV and
$\theta=93^\circ$ in the excitation energy range $E_x = 
7.6 - 11.7$ MeV. 
Bottom: the corresponding part of the $^{208}$Pb(p,p$'$) 
spectrum shown in 
Fig.~\protect\ref{fig:1}.}
\label{fig:1bis}
\end{figure}

Besides the analysis of the proton inelastic scattering data, 
we can also apply the entropy index method to the 
$^{208}$Pb(e,e$'$) data \cite{Kuh81}.
An experimental spectrum obtained for $E_e = 50$ MeV and
$\Theta = 93^\circ$ is presented in the upper
part of Fig.~\ref{fig:1bis}.
In the (e,e$'$) case, the high-resolution part of the 
experimental spectrum is limited to $E_x = 7.6 - 11.7$ MeV, 
the background is removed and 
the experimental resolution is again around 50 keV.                 
The part of the (p,p$'$) spectrum  corresponding to the
same energy interval is plotted in the lower part of 
Fig.~\ref{fig:1bis}.

\begin{figure}[h]
\begin{center}
%\epsfig{figure=eim-fig4.ps,
%height=6.0cm,angle=0}%,bbllx=1cm,bblly=3.5cm,bburx=19.5cm,bbury=26cm,clip=}
%\vspace{0.2cm}
\includegraphics*[height=6cm]{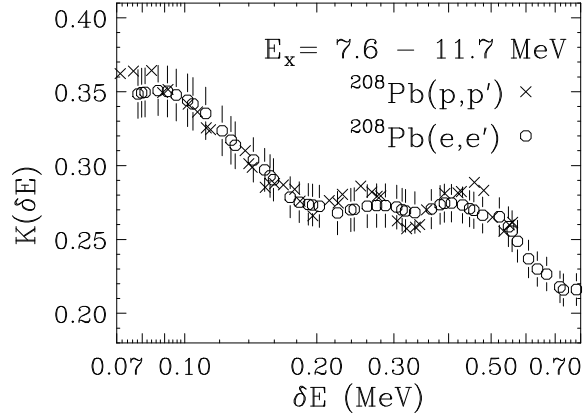}
\end{center}                                     
\caption{Comparison between the entropy index evolutions 
obtained 
from the $^{208}$Pb(p,p$'$) reaction (crosses) and the 
$^{208}$Pb(e,e$'$) reaction (circles). 
In both cases, the energy interval is $E_x = 7.6 - 11.7$ 
MeV. }
\label{fig:3}
\end{figure}

A detailed correspondence exists in the fine structure,
up to about 10 MeV even on a level-by-level basis \cite{Kam97}.
Similarities in the fine structure
observed between the two experiments
is indeed confirmed by the entropy index analysis.
Figure \ref{fig:3} presents the variation of $K(\delta E)$
as a function of $\delta E$ in the electron scattering case 
(circles) as well as for proton scattering (crosses).
The similarity in the fine structures observed in Fig.~\ref{fig:1bis} 
is reflected in a perfect superposition of the curves obtained from
these two independent experiments. 
This agreement provides further confidence in the estimated 
localization of curvature changes in $K\left( \delta E \right)$.  
 
\section{Comparison with second RPA results}

The interpretation of the fine-structure scales 
is far from being straightforward. Indeed, as we pointed out 
in the introduction, having precise information on 
the scales present in the damping of giant resonances is of particular
interest since it may help to understand which mechanisms are 
involved in the internal mixing. 
On the theory side, many physical effects might 
contribute to the damping \cite{Ber94}. 
Already at the mean-field level, 
the Landau fragmentation may introduce a typical scale. 
However, qualitative agreement with experiment can only be 
achieved when models include two-body effects 
as in extended RPA approaches \cite{Dro90,Kam97,Lac98}. 
In that case, we expect different effects due to the coupling 
of 1p-1h with 2p-2h states. 
The coherent coupling to low-lying collective surface vibrations
may introduce fragmentation\cite{Ber83}.
In addition, strong coupling may lead to a reduction of widths,
either from interferences due to
coupling through common decay channels \cite{Mah69,Sok89,Sok97-2} 
or due to motional narrowing \cite{Bro87}.
Furthermore, different assumptions are used in 
models in order to treat the decay channels of the collective 
states and/or the interaction matrix elements. 
For example, statistical assumptions might be needed
when the excitation energy increases\cite{Sok88,Fra96},
while in a microscopic picture like the SRPA, 
in particular at low excitation energy, only a few 2p-2h states 
are coupled to the collective states and statistical assumptions 
may break down. 

\begin{figure} [h]
\begin{center}
%\epsfig{figure=eim-fig6.ps,
%height=5.0cm,angle=0}%,bbllx=1cm,bblly=3.5cm,bburx=19.5cm,bbury=26cm,clip=}
%\vspace{0.2cm}
\includegraphics*[height=5cm]{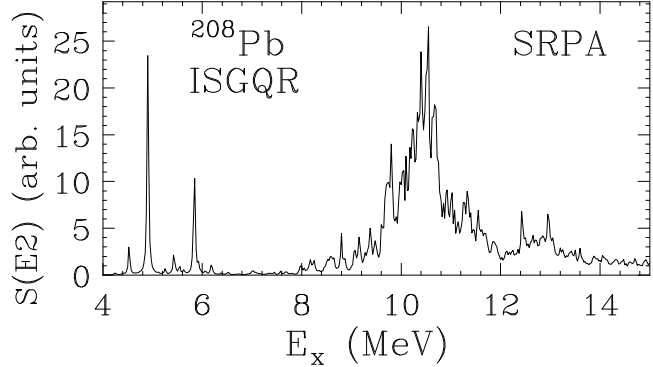}
\end{center}                                     
\caption{The SRPA isoscalar giant quadrupole strength 
function 
in $^{208}$Pb. 
The presented result was calculated with a width of the
complex selfenergy to give a resolution similar to the 
experimental
spectra in Figs.~\protect\ref{fig:1} and 
\protect\ref{fig:1bis}.}
\label{fig:5}  
\end{figure}

In this section, we apply the entropy index method to  
SRPA results for the isoscalar giant quadrupole response
in $^{208}$Pb. 
The calculation is based on the M3Y interaction \cite{Ber77}
with some adjustment of the short-range part which allows
to reproduce the experimental centroids of the low-multipolariy
electric giant resonances in $^{208}$Pb.
A truncation of the 2p-2h configuration space is necessary,
e.g., at the upper limit of the calculation $E_x = 20$ MeV
one would have to include about $1.5 \times 10^4$ 2p-2h states.
The method used here focuses on diagonal matrix elements
in the 2p-2h subspace.
Their distribution can be approximated by a Gaussian assuming
random fluctuations.
All configurations associated with matrix elements exceeding
this Gaussian fit are included in the further analysis
(about 3000 in the present example).
The complex SRPA selfenergy is chosen to attain a finite 
resolution similar to the experimental data.
The calculated strength function of the ISGQR is presented
in Fig.~\ref{fig:5}. 
At the RPA level (not shown) the strength function consists 
essentially of a single collective state around 12 MeV. 
By introducing 2p-2h components, the FWHM strongly increases
and fine structure appears on top of the global shape.  

\begin{figure}[h]
\begin{center}
%\epsfig{figure=eim-fig7.ps,
%height=5.0cm,angle=0}%,bbllx=1cm,bblly=3.5cm,bburx=19.5cm,bbury=26cm,clip=}
%\vspace{0.2cm}
\includegraphics*[height=5cm]{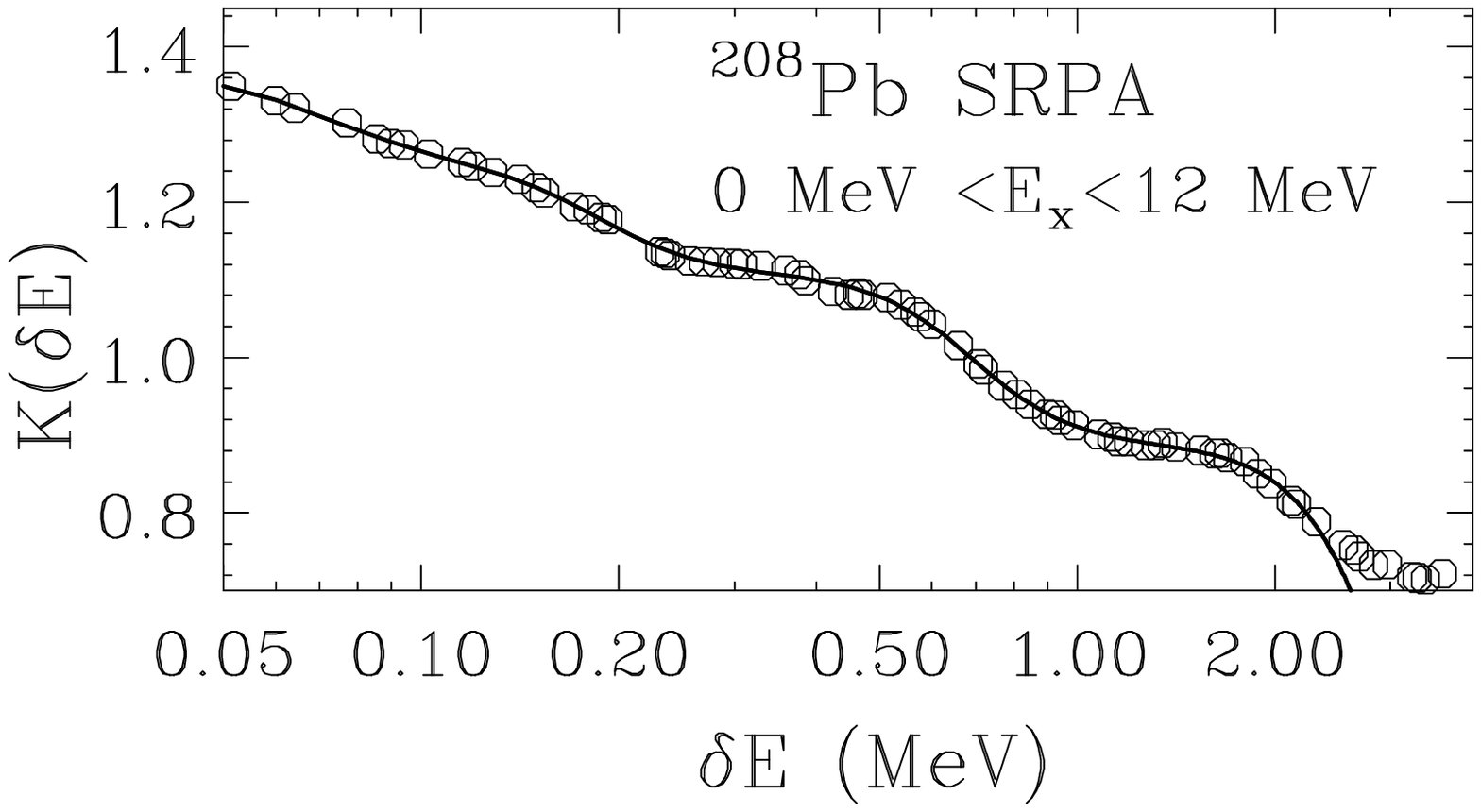}
\end{center}                                     
\caption{
The variation of the entropy index $K(\delta E)$ as a 
function
of the scale $\delta E$ obtained from the $^{208}$Pb SRPA 
strength function for $0$ MeV $<E_x < 12$ MeV. 
The solid line is a fit of expression 
(\protect\ref{eq:fit})
using a sum of four Fermi-Dirac functions.}
\label{fig:6}
\end{figure}

  The result obtained from the entropy index method analysis 
for this SRPA spectrum and $0 MeV < E_x < 12$ MeV 
is shown in Fig.~\ref{fig:6}.
The upper energy limit is set to exclude any modification
of the results due to the truncation of the 2p-2h model space.
Using the same fitting procedure as in the experimental case 
permits to identify scales of fluctuations at 2.1 MeV, 400 keV, 
and 120 keV. The two smaller values show very good agreement 
with the experimental observation while the first one
is somewhat larger.
This difference might result from the presence of multipolarities
other than $L = 2$ in the (p,p$'$) spectrum.
For example, the isoscalar giant monopole resonance centered
around 14 MeV is clearly visible in the data. 
The generally good agreement suggests that the origin of the
fine-structure scales is indeed 
due to the coupling of the 2p-2h excitations to the collective 
states and that their magnitude characterizes the two-body
components of the nuclear Hamiltonian, i.e.\ the coupling 
matrix elements and density of states.

\section{Conclusion}

In this paper, using the model-independent entropy index method
in high-resolution proton and electron elastic scattering, 
we show that multiple scales of the fine structure 
appear in the damping of the isoscalar giant quadrupole 
resonance of $^{208}$Pb. In addition to the entropy index method 
a fitting procedure has been developed that enables a more precise
estimate of the magnitude of these scales. 
Fine-structure scales appear at 1.1 MeV, 460 keV and 125 keV,
in perfect agreement between electron and proton scattering data. 
The application of the entropy index method to a 
SRPA strength function gives results in very satisfactory agreement with 
experiment while the RPA alone is not able to exhibit any of these scales. 
This provides a strong argument that the observed scales result from the 
decay of collective modes into 2p-2h states. 
In particular, they should be connected to the density 
of 2p-2h states and the coupling matrix elements of the in-medium 
residual interaction associated with 
this damping mechanism which is a rather unique way to 
infer properties of two-body components in the nuclear system. 
The quantitative knowledge of the scales 
may help to improve effective interactions which 
are generally fitted to mean-field properties only. 
In addition, the results argue in favor of a hierarchy of complexity in
the internal components of the nucleus: one-body, two-body, three-body... 
One may hope in the near future, with improved experiments, 
to uncover even smaller fine structure scales 
that are connected to more complex internal degrees of freedom
and go a step further in our understanding of the transition from
order to chaos in nuclear systems.

\section*{Acknowledgments}

One of us (D.~L.) gratefully acknowledges the Technische Universit\"at
of Darmstadt for the warm hospitality extended to him during his visit.
J.~Lisantti is thanked for providing us with spectra of the
$^{208}$Pb(p,p$'$) experiment.
This work was supported by the DFG under contract number Ri 242/12-2
and by the NSF grant NSFPHY98-00978.

\begin{thebibliography} {99}

\bibitem{BBB98} {P. F. Bortignon, A. Bracco, and R. A. Broglia,
{\it Giant Resonances: Nuclear Structure at Finite Temperature}
(Harwood Academic, Amsterdam, 1998).}
 
\bibitem{Row70}  {D.J. Rowe, {\it Nuclear Collective Motion}
(Methuen, London, 1970).}

\bibitem{Rin81}  {P. Ring and P. Schuck, {\it The Nuclear Many-Body Problem}
(Springer, New York, 1980).}

\bibitem{Dro90}  {S. Dro\.{z}d\.{z}, S. Nishizaki, J. Speth, and J. Wambach,
Phys. Rep. {\bf 197}, 1 (1990).}

\bibitem{Cas90}  {W. Cassing and U. Mosel, 
Prog. Part. Nucl. Phys. {\bf 25}, 235 (1990).}

\bibitem{Rei94}  {P.-G. Reinhard and C. Toepffer, 
Int. J. Mod. Phys. E {\bf 3}, 435 (1994).}

\bibitem{Abe96}  {Y. Abe, S. Ayik, P. G. Reinhard, and E. Suraud, 
Phys. Rep. {\bf 275}, 49 (1996).}

\bibitem{Zel96}  {V. G. Zelevinsky, 
Annu. Rev. Nucl. Part. Sci. {\bf 46}, 237 (1996).}

\bibitem{Sch75} {A. Schwierczinski, R. Frey, A. Richter, Spamer E., 
H. Thiessen, O. Titze, T. Walcher, S. Krewald, and R. Rosenfeld, 
Phys. Rev. Lett. {\bf 35}, 1255 (1975).}

\bibitem{Kuh81} {G. K\"uhner, D. Meuer, S. M\"uller, A. Richter,
E. Spamer, O. Titze, and W. Kn\"upfer, 
Phys. Lett. {\bf 104B}, 189 (1981).}

\bibitem{Kam97} {S. Kamerdzhiev, J. Lisantti, P. von Neumann-Cosel, A.
Richter, G. Tertychny, and J. Wambach, 
Phys. Rev. C {\bf 55}, 2101 (1997).}

\bibitem{Lis91}
J. Lisantti, E. J. Stephenson, A. D. Bacher, P. Li, R. Sawafta,
P. Schwandt, S. P. Wells, S. W. Wissink, W. Unkelbach, and
J. Wambach,
Phys. Rev. C {\bf 44}, R1233 (1991).

\bibitem{Win83} {J. Winchenbach, K. Pingel, G. Holzwarth, G. K\"uhner, and
A. Richter, 
Nucl. Phys. {\bf A410}, 237 (1983).}

\bibitem{Eri63}  {T. Ericson, 
Phys. Rev. Lett. {\bf 5}, 430 (1960); 
Ann. Phys. (N.Y.) {\bf 23}, 390 (1963).}

\bibitem{Kil87}  {G. Kilgus, G. K\"{u}hner, S. M\"{u}ller, A. Richter, 
and W. Kn\"{u}pfer, 
Z. Phys. A {\bf 326}, 41 (1987).}

\bibitem{Lac99} {D. Lacroix and Ph. Chomaz, 
Phys. Rev. C {\bf 60}, 064307 (1999).}

\bibitem{Hwa98}  {R.C. Hwa, physics/9809041.}

\bibitem{Ber94}  {G. F. Bertsch and R. A. Broglia {\it Oscillation in Finite
Quantum Systems} (Cambridge University, Cambridge, 1994).}

\bibitem{Lac98}  {D. Lacroix, Ph. Chomaz and S. Ayik, 
Phys. Rev. C {\bf 58}, 2154 (1998).}

\bibitem{Ber83}  {G. F. Bertsch, P. F. Bortignon and R. A. Broglia, 
Rev. Mod. Phys. {\bf 55}, 287 (1983).}

\bibitem{Mah69} {C. Mahaux and H. A. Weidenm\"uller, {\it Shell Model
Approach to Nuclear Reactions} (North Holland, Amsterdam, 1969).}

\bibitem{Sok89} {V. V. Sokolov and V. G. Zelevinsky, 
Nucl. Phys.  {\bf A504}, 562 (1989); 
Ann. Phys. {\bf 216}, 323 (1992).}

\bibitem{Sok97-2} {V. V. Sokolov, I. Rotter, D. V. Savin, and M. M\"uller, 
Phys. Rev. C {\bf C56}, 1031 (1997); 
{\it ibid} 1044.}

\bibitem{Bro87}  {R. A.  Broglia, T. D{\o}ssing, B. Lauritzen, 
and B. R. Mottelson, 
Phys. Rev. Lett. {\bf 58}, 326 (1987).}

\bibitem{Sok88} {V. V. Sokolov and V. G. Zelevinsky, 
Phys. Lett. B {\bf 202}, 10 (1988).}

\bibitem{Fra96} {N. Frazier, B. A. Brown and V. G. Zelevinsky, 
Phys. Rev. C {\bf 54}, 1665  (1996).}

\bibitem{Ber77}{G. F. Bertsch, J. Borysowicz, H. McManus, and W. G. Love,
Nucl. Phys. {\bf A284}, 399 (1977).}
 
\end{thebibliography}
\end{document}